\documentclass[fleqn,10pt]{wlscirep}
\usepackage[utf8]{inputenc}
\usepackage[T1]{fontenc}
\title{Dispersion Characterization and Pulse Prediction with Machine Learning}
\author[1]{Sanjaya Lohani}
\author[1]{Erin M. Knutson}
\author[1]{Wenlei Zhang}
\author[1,*]{Ryan T. Glasser}
\affil[1]{Tulane University, New Orleans, LA 70118, USA}

\affil[*]{rglasser@tulane.edu}

\begin{abstract}
In this work we demonstrate the efficacy of neural networks in the characterization of dispersive media.  We also develop a neural network to  make  predictions  for  input  probe  pulses  which propagate through a nonlinear dispersive medium, which may be applied to predicting optimal pulse shapes for a desired output.  The setup requires only a single pulse for the probe, providing considerable simplification of the current method of dispersion characterization that requires frequency scanning across the entirety of the gain and absorption features. We show that the trained networks are able to predict pulse profiles as well as dispersive features that are nearly identical to their experimental counterparts. We anticipate that the use of machine learning in conjunction with optical communication and sensing methods, both classical and quantum, can provide signal enhancement and experimental simplifications even in the face of highly complex, layered nonlinear light-matter interactions.
\end{abstract}
\begin{document}

\flushbottom
\maketitle
%
%
\thispagestyle{empty}

\section*{Introduction}

Optical pulse propagation through dispersive media often results in significant temporal distortion. The ability to characterize a medium based on its dispersive effects on pulses provides, for example, a route toward remote sensing of unknown media, and can provide information as to how to optimize an optical communications platform. Here we develop and experimentally implement neural networks with the ability to make predictions of input pulse shapes and dispersive features at the output of a four-wave mixing interaction in a warm gas of atoms; i.e. at the receiving end of an optical communications or remote sensing scheme. Four-wave mixing (FWM) in atomic vapor may be used to generate twin beams that have been shown to be useful in imaging \cite{boyer_entangled_2008, shi_quantum_2017}, spectroscopy \cite{thiel2008four}, and communications, both classical and quantum \cite{cai_quantum-network_2015}. Furthermore, the resultant two-mode squeezed light  \cite{mccormick_strong_2007} may be used to realize quantum steering \cite{wang_quantum_2017} and continuous-variable quantum teleportation \cite{Diao_theoretical_2019}, to improve on metrological limits \cite{hudelist_quantum_2014}, and to generate high-purity narrow-band single photons  \cite{macrae_tomography_2012}, among many other applications \cite{camacho_four-wave-mixing_2009, corzo_multi-spatial-mode_2011, turnbull_multi-spatial-mode_2014}. Some advantages FWM has over other methods of generating intensity-correlated beams are that there is no need for a cavity, and the bright output modes are generated such that they are spatially separated (as well as frequency-separated) from the pump beam(s) \cite{thiel2008four}. 
\begin{figure}[h!]
\centering
\includegraphics[width=0.65\linewidth]{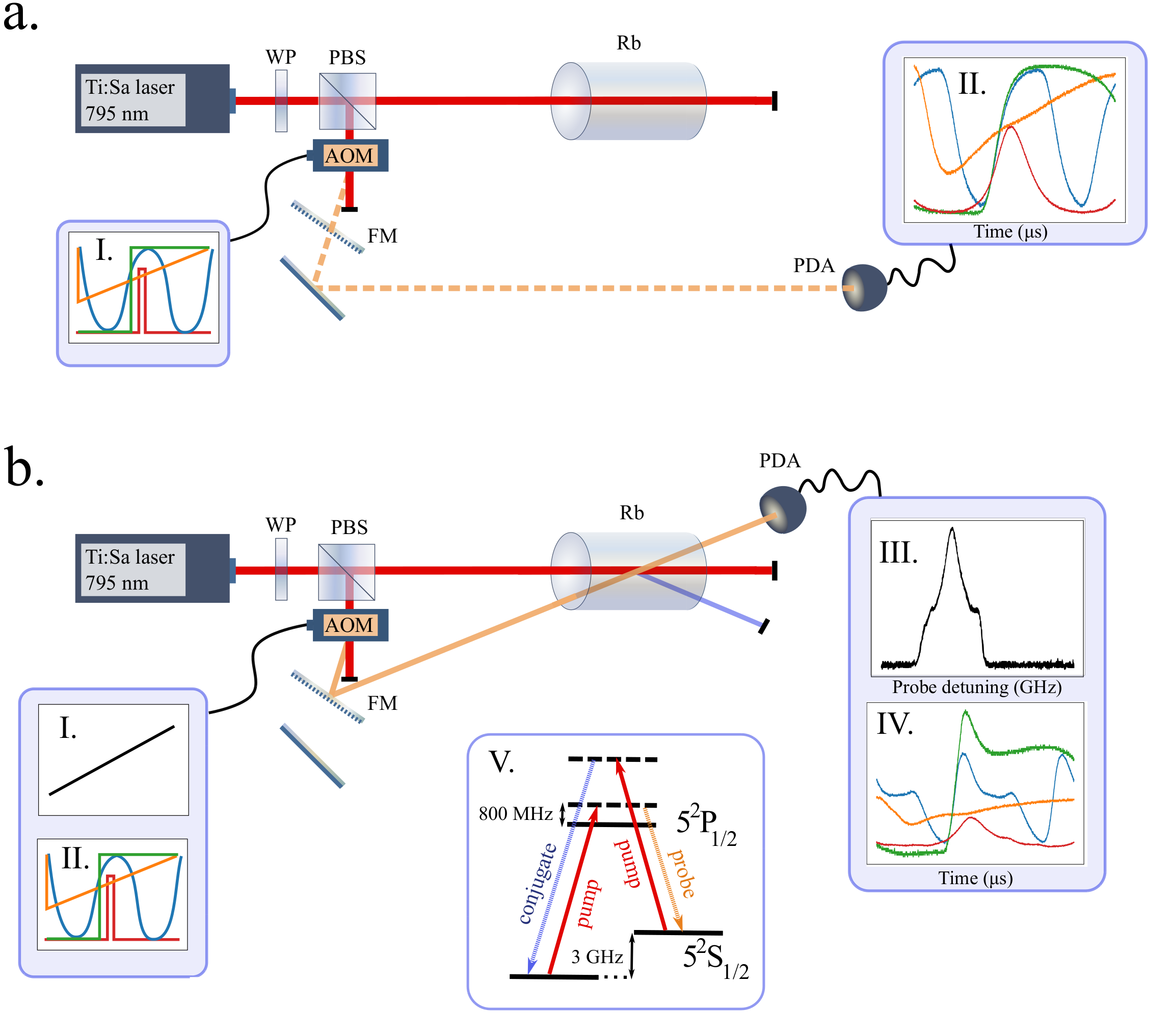}
\caption{(a.) Experimental setup for detecting reference (input) pulses. A Ti:Sapphire laser is locked to a wavelength of approximately 795 nm and passed through a $\lambda/2$ waveplate (WP) then split on a polarizing beam splitter (PBS). A portion of the light is passed through an acoustic-optical modulator (AOM), which is modulated with various pulses and waveforms (inset a-I) from an arbitrary waveform generator. The intensity-modulated and frequency-shifted light (orange) then bypasses a flip mirror (FM) and is incident on a detector (PDA), resulting in the reference probe pulses (a-II). 
(b.) Experimental setup for detecting gain lines and output probe pulses. The AOM is either scanned in frequency over $\sim$100 MHz (b-I) or pulsed as before (b-II), resulting in a gain line (b-III) or output probe pulses (b-IV) respectively. 
An energy level diagram for the FWM process is shown in inset b-V. }
\label{fig:Figure_1}
\end{figure}
The generally phase-insensitive modes also need not be spatially Gaussian or symmetric \cite{cao_experimental_2017, danaci_all-optical_2016, swaim_multi-mode_2018}, and the system is readily expanded either by cascading \cite{qin_experimental_2014, qin_experimental_2015} or adding more input beams \cite{wang_single-step_2017, liu_two-beam_2018, knutson_optimal_2018}. 

In many of these FWM applications, it is critical to measure the frequency response of the intensity of the ``probe,'' or seed, beam -- referred to as a gain line measurement -- in order to characterize the dispersion in the atomic medium, which depends in a complex manner on the atomic makeup, temperature, frequency of the beams involved, and so on. This is done by scanning the probe in frequency and detecting the resultant intensity while the pump is going through the medium, then subtracting the same measurement taken while the pump is blocked, in order to account for frequency dependence outside of the FWM interaction. In practice, this requires either a tunable probe laser or a frequency shifter such as an acoustic-optical modulator (AOM) or an electro-optical modulator (EOM). Accordingly, this measurement adds undesirable time and expensive equipment to systems that rely on four-wave mixing, or those that in general exhibit gain or absorption. 
It is often much simpler and faster, in practice, to pulse light than to scan it over a broad frequency range.  To address these issues, we introduce a scheme for predicting a portion of the gain line of an atomic FWM system using only single pulses with fixed center frequency on the probe mode, by measuring the distorted output pulses. In this way we take advantage of the fact that a short pulse is broad in its frequency spectrum, and no experimental frequency scanning or tuning is necessary. To do this, we train a convolutional neural network (CNN), which may also be used to correct for external dispersion or temporal distortion on the probe\cite{lohani_turbulence_2018} in its prediction of the gain line measurement, thereby correcting for the temporal distortion by optimizing the probe pulses at the input of the system. 

To generate FWM we pump a one-inch cell of rubidium vapor with 200 mW of 795 nm CW laser light. 
A $\approx$\,3\,GHz red-detuned, relative to the pump, probe beam crosses the pump at an angle of 0.8 degrees.  This probe beam may be scanned in frequency via an acousto-optic modulator (AOM) in order to generate gain line data in the standard frequency-scanning method, or amplitude modulated via the AOM (with a fixed frequency) and an arbitrary waveform generator in order to generate pulses, as shown in Fig. \ref{fig:Figure_1}.

\begin{figure}[h!]
\centering
\includegraphics[width=0.95\linewidth]{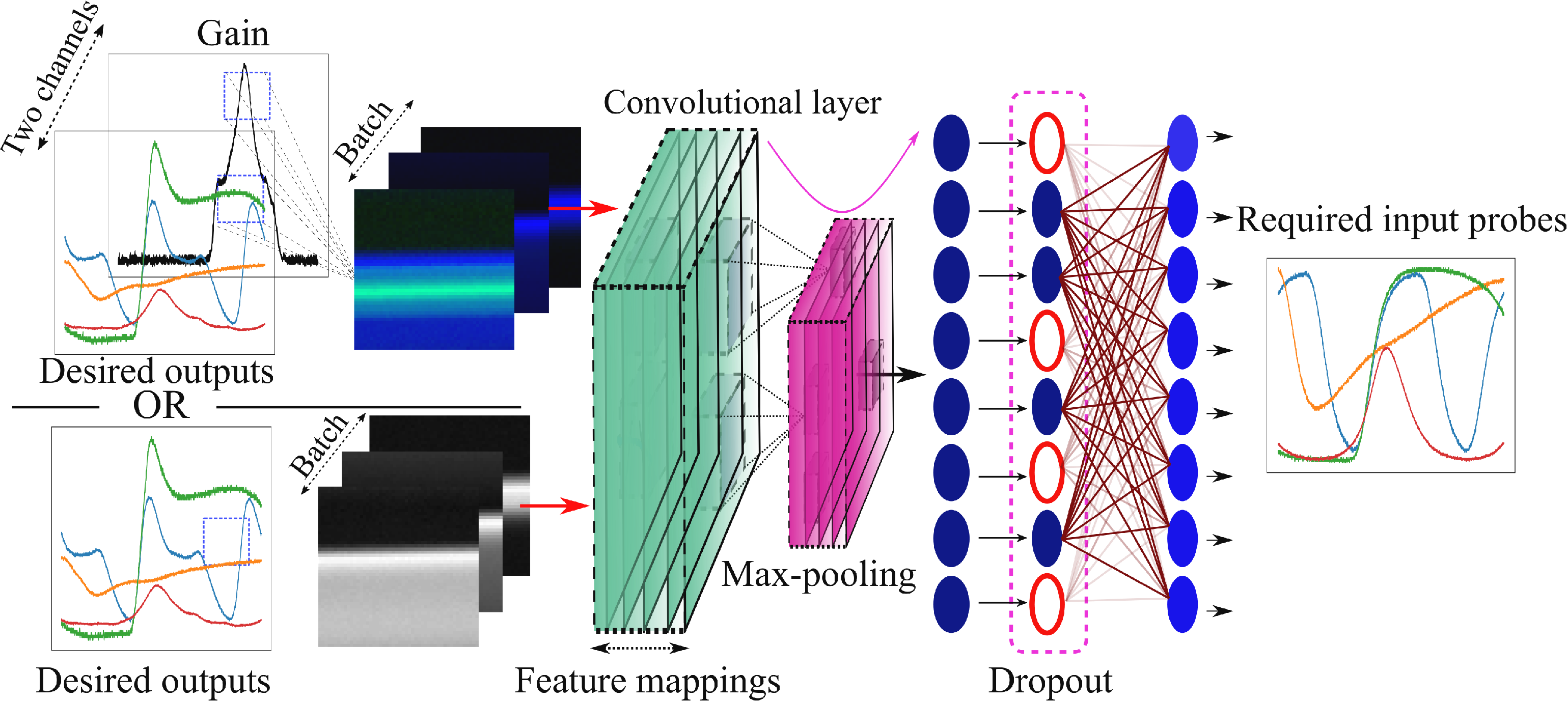}
\caption{Architecture of the neural network for unknown probe prediction using gain and desired outputs as the two channel input, or outputs only (without gain) as the single channel input.  Here the measured output probe and dispersion profiles are used to predict the input probe pulse. As discussed in the text the scheme is easily altered to predict the dispersion profile using the input and output probe pulses as the two channel input, or the output pulses only (without input pulses) as the single channel input.}
\label{fig:Figure_2}
\end{figure}

Machine learning techniques have been applied to various scientific and research fields \cite{lv_machine_2019,karpatne_machine_2018-7,deo_machine_2015,butler_machine_2018-3,hegde_use_2017-3,vu_shared_2018,tranter_multiparameter_2018-3,zahavy_deep_2018-1}, including using CNNs in the context of optical communications \cite{lohani_use_2018,tanimura_convolutional_2019-4,rahmani_multimode_2018,doster_machine_2017}. Additionally, deep neural networks have been shown to be useful in a variety of regression type optimization scenarios \cite{lotfinejad_comparative_2018,ye_modeling_2018-4,xu_regression_2015,tan_convolutional_2017}. 
Here we use a CNN to make predictions for the input probe pulses which propagate through a nonlinear dispersive medium, which often contains complex gain and absorption features. Additionally, this technique is able to predict the profile of unknown input pulses sent through a dispersive medium by using the measured output pulse profiles.  This flexibility allows for the developed system to be used in a variety of applications, including remote sensing of unknown materials, and for the optimization of optical pulse propagation through unknown media.

\section*{Results}
\begin{figure}[b!]
\centering
\includegraphics[width=0.85\linewidth]{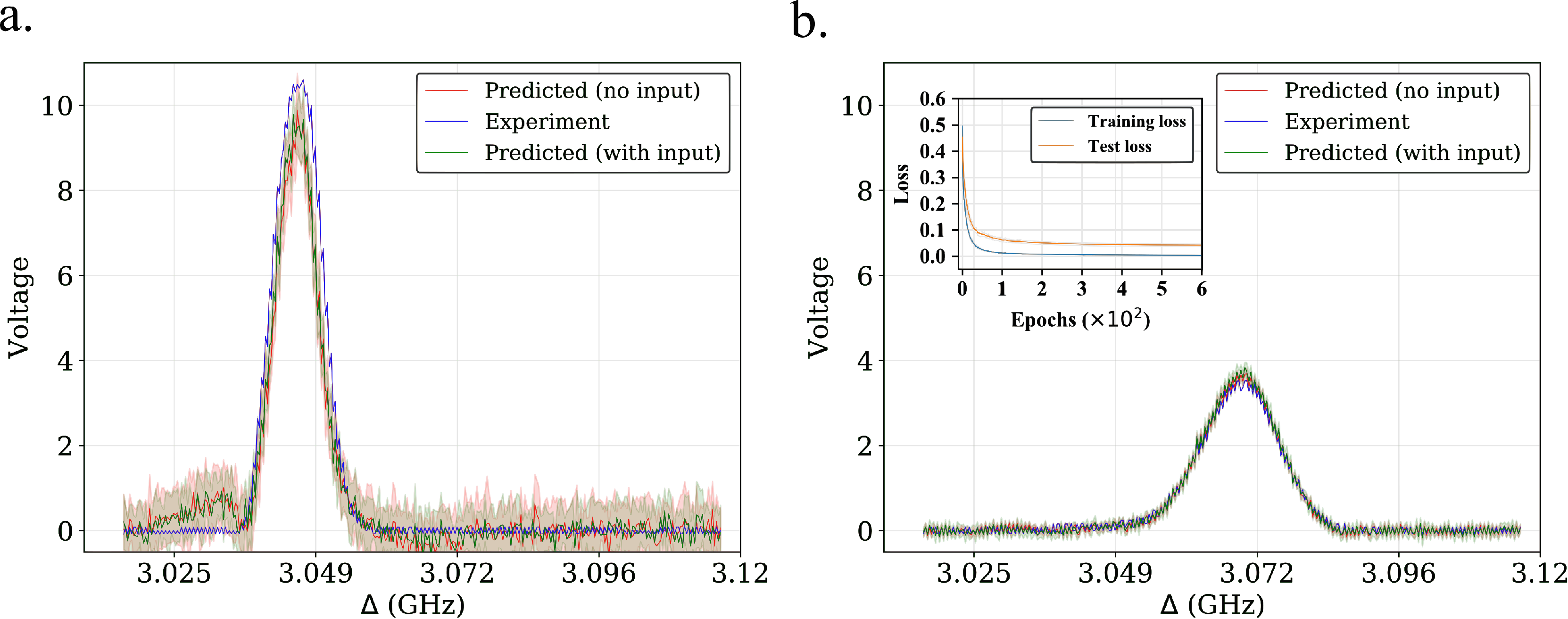}
\caption{Predicting the gain curve of a non-linear medium peaked approximately at a detuning of (a) 3.047 GHz, and (b) 3.071 GHz using FWM output probes and input probes as two channel training inputs (green), and FWM output probes only as single channel training inputs (red) to the network. The mean square loss at each epoch is shown in inset of (b).}
\label{fig:Figure_4}
\end{figure}

\begin{figure}[h!]
\centering
\includegraphics[width=0.7\linewidth]{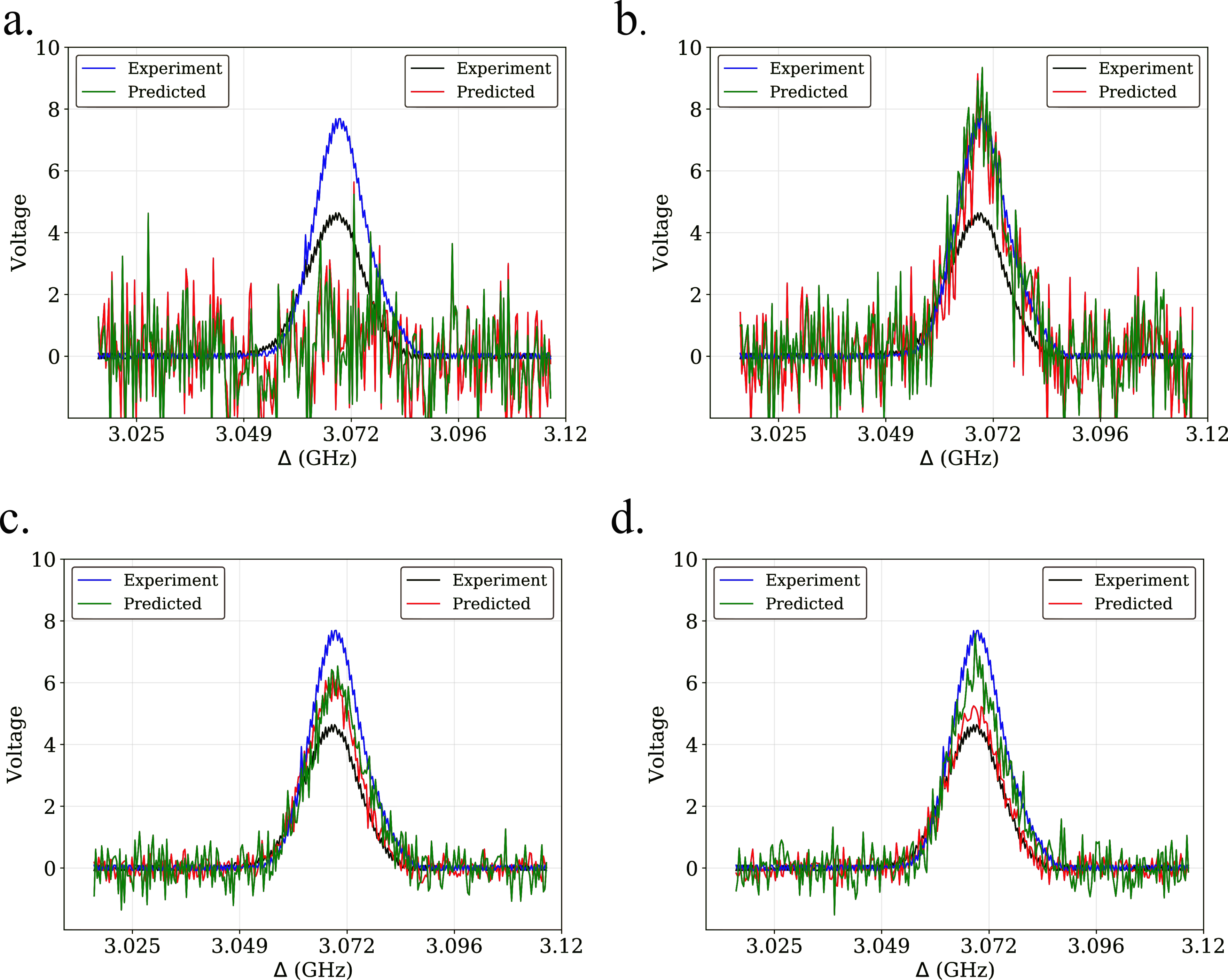}
\caption{Gain curve predictions using the training data with a set of (a) 2, (b) 4, (c) 8, and (d) 16 pulses, respectively.}
\label{fig:Figure_5}
\end{figure}

The CNN contains a single two dimensional convolutional layer with a kernel of size \Big[5, 5, 2\Big] (\Big[3, 3, 2\Big] for the results shown in Fig. \ref{fig:Figure_5}), where 2 represents the two-channel input. The convolutional layer has a stride length of 1 (2 for the results shown in Fig. \ref{fig:Figure_5}) and a rectified linear unit (ReLU) activation convolutes the input pulse (image) with a size of \Big[50, 50, 2\Big] to a size of \Big[46, 46, 10\Big], where 10 represents the number of feature mappings. Then we apply a zero padding such that the dimension of the image after the convolution, again, becomes \Big[50, 50, 10\Big]. After this, we apply a two dimensional max pool layer with a kernel of size \Big[2, 2\Big] that reduces the width and height of image to half its value, \Big[25, 25, 10\Big]. Next, we attach a fully connected layer (FCL) with 5,000 neurons (2,500 neurons for the single channel input case) to the output of the max-pooling followed by the ReLU activation function. Then we apply a dropout with a rate of $50\%$ to the outputs of the FCL. Finally, we connect the output of the FCL to an output layer consisting of 2,500 neurons (300 for gain curve predictions) followed by a linear activation function. Note that the hyperparameters of the neworks are manually optimized as discussed in \cite{lohani_use_2018}. In order to generate the two channel data set for predicting input pulse profiles, we stack the FWM output probes (desired outputs) and corresponding gain lines on each other. These are then randomly split into a training set and test set. 
The training set is then fed into the CNN, which makes predictions for the required input probe pulses to be sent through the Rb cell.
Examples of the desired outputs (FWM output probes), gains and corresponding required input probe pulses are shown in Fig. \ref{fig:Figure_2}, along with a schematic of the neural network architecture. This process is repeated many times with different initialization points for the given unknown test set of output probes and gains, and the required input probes are predicted and compared to the experimental input probes. Additionally, the CNN makes predictions for the input probes using only output probes as single-channel data (i.e no gain lines). Finally, we alter the system to make predictions for gain line profiles by using output probes and input probes stacked on each other as the two channel data set, as well as using only output probes as the single channel set. The predicted gains are again compared to the experimental values, as shown in Figs. \ref{fig:Figure_4} and \ref{fig:Figure_5}. For latter case, we benchmark how closely the predicted dispersion profiles fit the experimental data as the training data set is varied.

\begin{figure}[h!]
\centering
\includegraphics[width=0.7\linewidth]{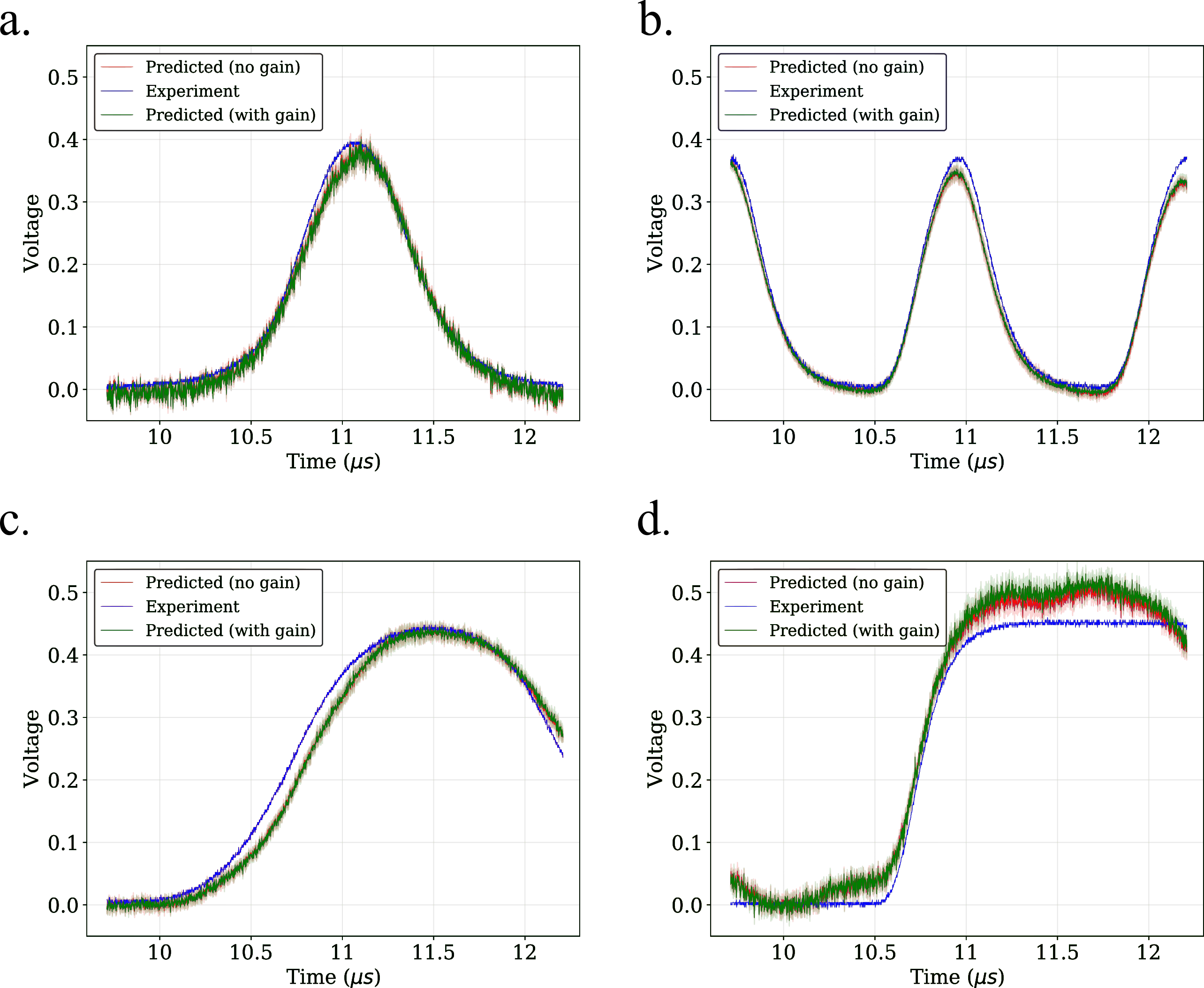}
\caption{(a-d) Input probe (pulse) predictions using, green: FWM output probes and gain profiles as two channels training inputs, and red: FWM output probes only (without gain) as a single channel data to train the network.}
\label{fig:Figure_3}
\end{figure}
In order to make predictions for input probe pulses by making use of a gain curve, we use two channels of data in the convolutional networks. Here the gain, input probe pulses, and output probes after FWM have 2,500 points between the time-scale of 9.71 $\mu s$ to 12.2 $\mu s$. First we convert these 2,500 points to a corresponding image of size $50\times50$. As a result we have a total of 64 different sets (images) of gain, output probes, and their corresponding input probes. Note that here the gain curve remains the same for all the combinations of output probes and their respective input probes. Next we randomly split them into training data consisting of 60 sets of pulses and testing data with 4 sets of pulses. Note that each training and testing set has an output probe stacked with a gain line so as to make the two channel data to the network, with the corresponding input probe as the target. 
The two channel data is scaled to have zero mean and unity variance before being fed into the CNN (no scaling is performed on the target probe pulses). After this, the network is trained with a learning hyper-parameter of 0.008 for up to 600 epochs using a stochastic batch optimization technique using adamoptimizer of tensorflow\cite{tensorflow2015-whitepaper}. Then we feed the unknown 4 sets of pulses (output probes and gain lines) to the pre-trained network to make the predictions for their corresponding input probes. The predicted results (green) versus experimentally measured (blue) pulses are shown in Fig. \ref{fig:Figure_3} (a-d). Similarly, we use only the FWM output probe as a single channel data (with no gain line) to the network and the corresponding input probe as the output of the network to train the network. With the same hyper-parameter settings as before, predictions made by the pre-trained network are shown by the red curves in Fig. \ref{fig:Figure_3} (a-d). The translucent bands, shaded green around the predicted green curves and shaded red around the predicted red curves represent one standard deviation from the mean value of 15 different trials.  This exact system may then be used to predict input pulse profiles for given desired output profiles, by using a desired output pulse with the gain line as the two channel input into  the network.

We now turn to predicting different gain curves using input probes and their corresponding FWM output probes. Note that here the input probe remains the same for all the different combinations of gain and corresponding output probe sets. We use 27 different combinations of gain lines and FWM output probes, which are randomly split into training data and testing data with 25 pulse sets and 2 pulse sets respectively. The input probes and FWM output probes are stacked on each other to make two channel data input to the network, with the corresponding gain line as the target. Note that we clip the gain curves to 300 points, corresponding to a 3.017 GHz to 3.117 GHz detuning, which is equal to the number of output neurons of the network, with the input and FWM output probes again consisting of 2,500 points. The CNN is trained with the same hyper-parameter settings (except now a learning hyper-parameter of 0.009) as described in previous paragraphs and makes predictions for unknown gain lines. We find the predicted gains (green curves) are nearly identical to the experimental values (blue curves) as shown in Fig. \ref{fig:Figure_4}. Similarly, we train the same network with only a FWM output probe (no input probes) as single channel data to the network and again make  predictions for the unknown gain lines. We again find significant overlap between the prediction results (red curves) and the experimental data. The predicted and experimental gains peaked approximately at a detuning of 3.047 GHz and 3.071 GHz, and are shown in Fig. \ref{fig:Figure_4} (a), and Fig. \ref{fig:Figure_4} (b), respectively. The translucent bands again represent one standard deviation from the mean value of 15 different trials. Furthermore, in the case of two channel input to the network, the mean square loss between the unknown target gain line and predicted gain at each epoch is shown in inset of Fig. \ref{fig:Figure_4} (b), which shows the loss is saturating after 200 epochs.  

Lastly, we investigate the improvement in making predictions of gain with respect to the number of training sets. In order to generate a prediction benchmark, we use 18 different gain curves peaked between 3.068 GHz to 3.074 GHz as the unknowns to be predicted, and vary the number of pulses used in the training data set.
We use network layers as discussed above with a learning hyper-parameter of 0.009. First we randomly choose 2 pulses out of 18 as the test data (unknown) and keep them fixed. After this we again randomly select 2, 4, 8, and 16 pulses as the training data each out of the remaining 16 pulses and then train the networks separately with them. Finally, the pre-trained networks make predictions for the unknown gains. As expected, we find the better gain predictions when using a higher number of training pulse sets as shown in Fig. \ref{fig:Figure_5}. Note that the gain predictions shown by the red and green curves correspond to the experimental gains shown by the black and blue curves, respectively. The unknown test gain curve predictions using training data with sets of 2, 4, 8, and 16 pulses are shown in Fig. \ref{fig:Figure_5} (a-d).      

\section*{Conclusion}
In conclusion, we have implemented convolutional neural networks with the ability to make estimations of unknown input pulses that have experienced distortion when passing through a dispersive atomic medium (nonlinear four-wave mixing in rubidium vapor), given the resultant distorted output pulses. We demonstrate that the predicted input probe pulse shapes and amplitudes match well with their experimental counterparts. In addition to straightforward classification, this method may be expanded as in an end-to-end communication or remote sensing system, to make predictions at the receiving end for completely unknown transmitted pulses propagating through different dispersive media. 
Once pre-trained, the networks may also directly be used to optimize the input pulses that should be sent through a dispersive medium, given a desired output (or received) pulse.
Additionally, with the same networks, we have demonstrated the successful prediction of gain lines -- a measurement over a range of probe frequencies -- using probe pulses with a single center frequency, thus requiring no scanning. This could considerably simplify experiments wherein it is important to characterize the approximate response of a medium to various frequency inputs, but where frequency scanning the relevant beam is difficult, time-consuming, or costly. 

\bibliography{Pulse}

\section*{Acknowledgements }

We acknowledge funding from the U.S. Office of Naval Research under grant number N000141912374, the National Science Foundation Graduate Research Fellowship under grant number DGE-1154145, as well as from Northrop Grumman - NG NEXT. This research was supported in part using high performance computing (HPC) resources and services provided by Technology Services at Tulane University, New Orleans, LA. 

\section*{Author contributions statement}
S.L. developed and implemented the neural networks, E.M.K and W.Z. performed the experiments, and R.T.G. developed and oversaw the project. All authors contributed to analyzing the data and writing the manuscript. 

\end{document}